\documentclass[aps,prl,twocolumn,superscriptaddress,amssymb,amscd,amsmath,groupedaddress,a4paper,showpacs]{revtex4-1} 
\usepackage{graphicx}
\usepackage{dcolumn}
\usepackage{bm}

\usepackage{color}
\usepackage{suffix}
\usepackage{mathtools}

\DeclarePairedDelimiterX\MeijerM[3]{\lparen}{\rparen}%
{\begin{smallmatrix}#1 \\ #2\end{smallmatrix}\delimsize\vert\,#3}

\newcommand\MeijerG[8][]{%
  G^{\,#2,#3}_{#4,#5}\MeijerM[#1]{#6}{#7}{#8}}

\WithSuffix\newcommand\MeijerG*[7]{%
  G^{\,#1,#2}_{#3,#4}\MeijerM*{#5}{#6}{#7}}

\begin{document}

%\preprint{}

\title{Quantum breaking of ergodicity in semi-classical charge transfer dynamics}

\author{Igor Goychuk}
 \email{igoychuk@uni-potsdam.de}
 
\affiliation{Institute for Physics and Astronomy, University of Potsdam, 
Karl-Liebknecht-Str. 24/25, 14476 Potsdam-Golm, Germany}

\date{\today}

\begin{abstract}
Does electron transfer (ET) kinetics within a single-electron trajectory description 
always coincide with the ensemble description? This fundamental 
question of ergodic behavior is scrutinized within a very basic semi-classical 
curve-crossing problem of quantum 
Landau-Zener tunneling between two electronic states with overdamped classical 
reaction coordinate.
It is shown that in the
limit of non-adiabatic electron 
transfer (weak tunneling) well-described by the Marcus-Levich-Dogonadze (MLD) 
rate the answer
is yes. However, in the limit of the so-called solvent-controlled adiabatic electron 
transfer a profound breaking of ergodicity occurs. The ensemble survival
probability remains
nearly exponential with the inverse rate given by the sum of the adiabatic curve crossing
(Kramers) time  
and inverse MLD rate. However, near
to adiabatic regime, the single-electron survival probability is clearly non-exponential
but possesses an exponential tail which agrees well with the ensemble description. Paradoxically,
the mean transfer time in this classical on the ensemble level 
regime is well described by the inverse of nonadiabatic quantum tunneling 
rate on a single particle level.

\end{abstract}

\maketitle

\begin{figure}
\vspace{1cm}
\resizebox{0.8\columnwidth}{!}{\includegraphics{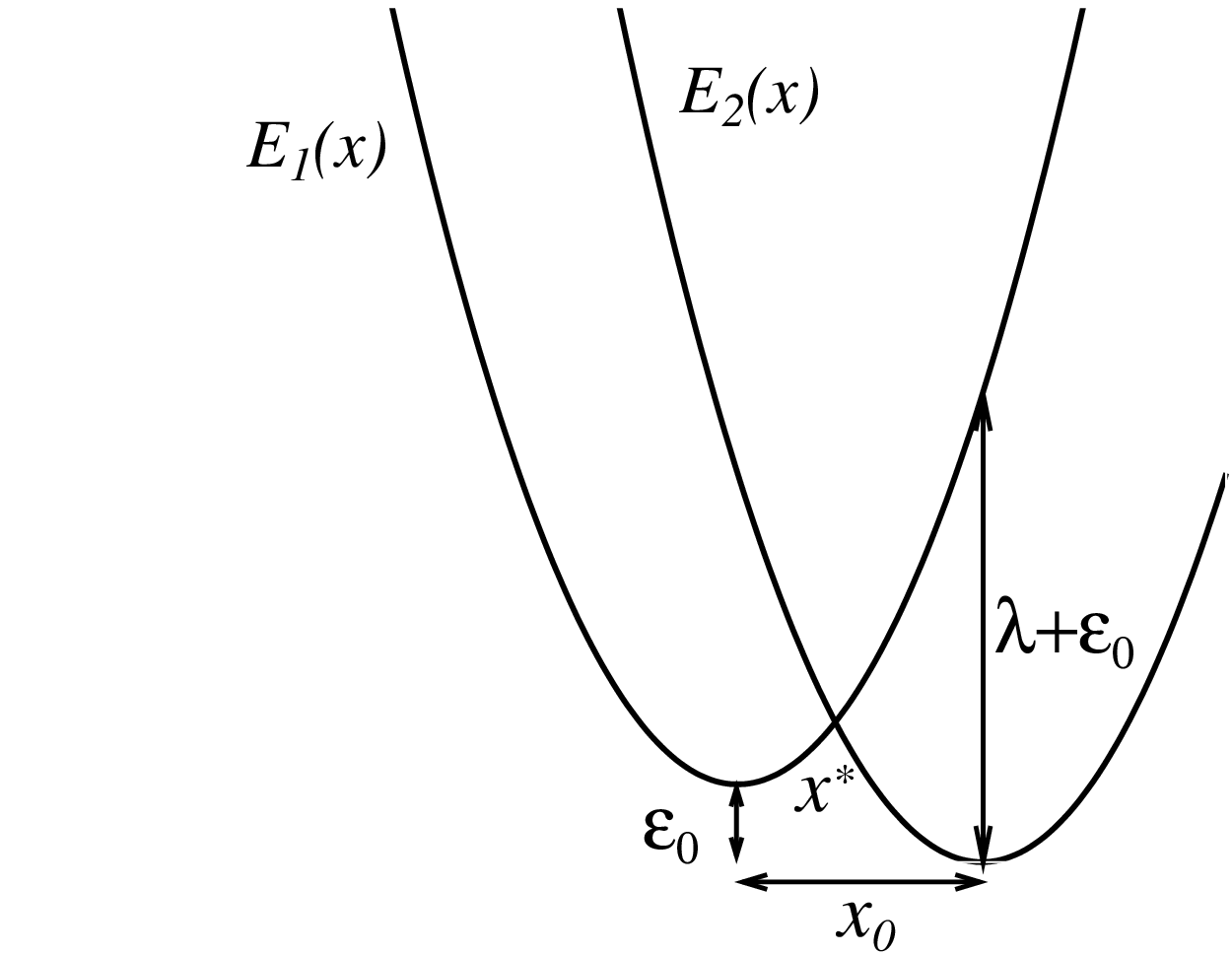}}
\caption{Curve crossing problem in the case of
two equal potential curvatures $\kappa$ (i.e. no nuclear frequency change occurs at electronic 
transitions). 
Diabatic electron energy levels $E_{1,2}(x)$ provide harmonic
potentials 
for the nuclear or molecular reaction coordinate $x$. $x_0$ is the nuclear equilibrium shift
for different electron energies, and $\epsilon_0$ is the corresponding electron energy
difference. $\lambda=\kappa x_0^2/2$  is nuclear (molecular) reorganization energy.}
\label{Fig1}       
\end{figure}

Discovery of ergodicity breaking on the level of single molecular stochastic dynamics 
\cite{Barkai} calls for re-examination of the basic models of stochastic transport in condensed
matter. Even some standard models like diffusion in Gaussian disordered potentials with short-range
correlations \cite{Zwanzig,HTB90} can be mesoscopically non-ergodic \cite{PRL14}. This work 
discovers ergodicity
breaking in another very popular and basic transport model based on
a curve-crossing tunneling problem \cite{Nitzan}. 
It  is fundamental for quantum transport in condensed matter with 
a famous Landau-Zener-St\"uckelberg (LZS) result for the probability of quantum transitions 
\cite{LZS,Nitzan}
\begin{eqnarray}\label{LZS}
P_{\rm LZ}(v)=1-\exp\left [- f(v)\right ]
\end{eqnarray}
between two diabatic quantum states $|1\rangle$ and $|2\rangle$, 
presenting a milestone. Here,
\begin{eqnarray}\label{Landau}
f(v)=\frac{2\pi}{\hbar}\frac{|V_{\rm tun}|^2}{|(\partial \Delta E(x))/\partial x)v|_{x=x_*}},
\end{eqnarray}
$v=\dot x$, is the result of the lowest second order quantum 
perturbation theory 
in the tunnel coupling $V_{\rm tun}$. It follows from the Fermi Golden Rule quantum transition rate
\begin{eqnarray}\label{GoldenRule}
\Gamma(x)=\frac{2\pi}{\hbar}|V_{\rm tun}|^2\delta(\Delta E(x))
\end{eqnarray}
applied at the level crossing point $x^*$, $\Delta E(x^*)=0$.
$\Delta E(x)=E_1(x)-E_2(x)$ is the difference of the 
diabatic energy levels, which depends on time via a modulation parameter $x(t)$, $\delta(x)$
is the Dirac's delta-function. Quantum system is characterized by the Hamiltonian
$\hat H(x)=E_1(x)|1\rangle \langle 1|+E_2(x)|2\rangle \langle 2|+
V_{\rm tun}(|1\rangle \langle 2|+|2\rangle \langle 1|)$, and 
the parameter $x(t)$ here is the
nuclear coordinate, see in Fig. \ref{Fig1}, 
which is treated classically (a mixed quantum-classical description), 
and $|i\rangle$, $i=1,2$ are two localized electronic states
between which electron can tunnel.
Within the Born-Oppenheimer approximation, $E_{1,2}(x)$ present
diabatic energy potentials for  $x(t)$.
Furthermore, within the harmonic approximation and assuming that no nuclear
frequency change occurs at electronic transitions,
$E_{i}(x)=\kappa (x-x_0\delta_{2,i})^2/2-\epsilon_0\delta_{2,i}$. Then, 
$\Delta E(x)=\epsilon_0-\lambda+2\lambda x/x_0$, where
$\epsilon_0$ is the difference of electron energy levels for equilibrium nuclei,
$x_0$ is the reaction coordinate shift, % at quantum transition, 
and $\lambda=\kappa x_0^2/2$ is  nuclear reorganization energy. Notice that 
electron tunnel distance has anything in common with $x_0$. Electron tunnels in space
once the transition $|1\rangle\to |2\rangle$, or $|2\rangle\to |1\rangle$ takes place. 
Likewise, blinking of a quantum dot occurs once it is in the light emitting
 quantum state. Depending on the coupling
strength $V_{\rm tun}$ 
and the velocity $v=\dot x$ at
the the crossing point $P_{\rm LZ}(v)$ can vary from 
$P_{\rm LZ}(v)\approx f(v) \propto |V_{\rm tun}|^2/|v|$
(nonadiabatic transition) to one (adiabatic transition).

Within a classical treatment of the reaction coordinate $x$, one considers it as a particle
of mass $M$ subjected to viscous frictional force $\eta v$, with a friction coefficient $\eta$,
and zero-mean white Gaussian thermal noise of the environment $\xi(t)$ at temperature $T$. The 
friction and noise are related by the fluctuation-dissipation relation 
$\langle \xi(t)\xi(t')\rangle =2k_BT\eta\delta(t-t')$, where $\langle ...\rangle$ denotes ensemble
averaging. Stochastic dynamics of $x$ follows Langevin equation
\begin{eqnarray}\label{Langevin}
M\ddot x+\eta \dot x +\frac{\partial E_{i}(x)}{\partial x}=\xi(t),
\end{eqnarray}
which depends on the quantum state $|i\rangle$. The electron-reaction
coordinate dynamics can be described in a semi-classical approximation by a mixed
quantum-classical dynamics of the reduced density matrix $\rho_{ij}(x,v,t)$, where the quantum
degree follows quantum dynamics while the dynamics in $(x,v)$ phase space for a fixed
quantum state $i$ is classical. Generally, it is described by the Kramers-Fokker-Planck equation (KFPE). 
In the overdamped case, $\eta\gg\sqrt{M\kappa}$, the reaction coordinate velocity
is thermally distributed, $P_M(v)= \exp[-v^2/(2v_T^2)]/\sqrt{2\pi v_T^2}$, $v_T=\sqrt{k_BT/M}$, all the time.
In a singular limit of  $M\to 0$, 
KFPE for a fixed state $i$  
reduces to Smoluchowski-Fokker-Planck dynamics,
$\dot p_{i}(x,t)=\hat L_{i} p_{i}(x,t)$ characterized by the Smoluchowski  operator
$\hat L_{i}=D(\partial /\partial x)\{\exp[-\beta E_{i}(x)](\partial /\partial x)
\exp[\beta E_{i}(x)]\}$. Here, $\beta=1/k_BT$ is inverse temperature, and
$D=k_BT/\eta$ is diffusion coefficient. The corresponding
semi-classical description is well known under the label of Zusman-Alexandrov equations 
\cite{Zusman,Garg}.
Within it,  the dynamics of populations $p_{i}(x,t):=\int \rho_{ii}(x,v,t)dv$ is described by
\begin{eqnarray}\label{Zusman}
\dot p_1(x,t)&=&-K(x)[ p_1(x,t)-p_2(x,t)]+\hat L_1 p_{1}(x,t), \nonumber \\ 
\dot p_2(x,t)&=&K(x)[ p_1(x,t)-p_2(x,t)]+\hat L_2 p_{2}(x,t),
\end{eqnarray}
after excluding (projecting out) the dynamics of quantum coherences. 
Here, $K(x)$ is a complicated expression \cite{JCP00} which in the so-called contact approximation
is simply $K(x)\approx \Gamma(x)$ \cite{Zusman},  where $\Gamma(x)$ is 
the Golden Rule expression in (\ref{GoldenRule}).
Indeed, for a strong electron-nuclear coupling ($\lambda\gg V_{\rm tun}$) 
and in the limit where the quantum effects in the reaction coordinate dynamics
are entirely neglected, this approximation is well justified \cite{Zusman,Garg}.
It presents a very important reference point,
 which allows also for further generalizations toward anomalous subdiffusive dynamics
 of the reaction coordinate \cite{TangMarcus}.  
 Indeed, within this approximation one obtains very
 elegant and important analytical results.  Consider first
 very small $V_{\rm tun}$, with the reaction coordinated being thermally equilibrated,
 $P^{(eq)}_i(x)=\exp[-(x-x_0\delta_{2,i})^2/(2x_T^2)]/\sqrt{2\pi x_T^2}$, where
  $x_T=\sqrt{k_BT/\kappa}=x_0\sqrt{k_BT/(2\lambda)}$ is thermal width, before each and every 
 quantum
 transition occurs. Then, the nonadiabatic quantum transition rate is
 \begin{eqnarray}\label{Marcus}
 k_{ i}^{\rm (nad)} = \int_{-\infty}^{\infty}P^{(eq)}_i(x)\Gamma(x)dx %\nonumber \\
  = \frac{2\pi V_{\rm tun}^2 }{\hbar\sqrt{\pi\lambda k_BT}}
  e^{- \frac{E^{(a)}_{i}}{k_BT}}
 \end{eqnarray}
 with activation energies $E^{(a)}_{1,2}=(\epsilon_0\mp\lambda)^2/(4\lambda)$. This is celebrated
 Marcus-Levich-Dogonadze formula \cite{Marcus,Levich,Nitzan}.
 Parabolic dependence of $E_{i}^{(a)}$ on 
 $\epsilon_0$ is famously
 known as Marcus parabola. Notice in this respect that the so-called inverted regime of electron
 transfer for $\epsilon_0>\lambda$ is entirely quantum-mechanical feature which is physically
 impossible within an adiabatic classical treatment.
 
 With the increase of $V_{\rm tun}$ the reaction coordinate dynamics becomes ever more important
 and it can limit the overall rate. The following expression has been derived \cite{JCP00} 
 from Eq. (\ref{Zusman})
 \begin{eqnarray}\label{analytic1}
 k_{ i} = \frac{k_{ i}^{\rm (nad)}}{1+\tau^{\rm (ad)}_1k_{ 1}^{\rm (nad)}+
 \tau^{\rm (ad)}_2k_{ 2}^{\rm (nad)}},
 \end{eqnarray}
 where
 \begin{eqnarray}\label{analytic1a}
 \tau^{\rm (ad)}_i=\tau\Bigg (\ln(2)+
 2\Big ( \frac{E_i^{(a)}}{k_BT} \Big ) \sideset{_2}{_2}{\mathop{F}}\Big (
 1,1;\frac{3}{2},2;\frac{E_i^{(a)}}{k_BT} \Big) \Bigg )
 \end{eqnarray}
 is the mean escape time in the parabolic potential with cusp, and $\tau=\eta/\kappa$
 is the reaction coordinate relaxation time. 
 Here, $\sideset{_2}{_2}{\mathop{F}}(a,b;c,d;z)$ is a generalized hypergeometric
series \cite{Gradstein}. For $E_i^{(a)}\gg k_BT$, $\tau^{\rm (ad)}_i\approx 
\tau \sqrt{\frac{\pi k_BT}{E_a^{\pm}}}
\exp{\Big ( \frac{E_i^{(a)}}{k_BT} \Big) }$ \cite{Zusman}. Hence for large activation barriers
and $\tau^{\rm (ad)}_ik_{ i}^{\rm (nad)}\gg 1$, 
\begin{eqnarray}
k_{ i}\approx  k_{ i}^{\rm (ad)} = 
 \frac{1}{\tau}
\sqrt{\frac{E_1^{(a)}E_2^{(a)}}{\pi \lambda k_BT}}
e^{-\frac{E_i^{(a)}}{k_BT}} \;,
\end{eqnarray}
which is adiabatic Marcus rate.
 For a particular case $\epsilon_0=0$, $k_{ 1,2}^{\rm (ad)}$ coincides with the
 Kramers rate for the adiabatic transitions in the cusp potential consisting
 of two pieces of diabatic curves in Fig. \ref{Fig1} \cite{HTB90}. Hence, for a sufficiently 
 large $V_{\rm tun}$
 ET becomes classical and adiabatic within this ensemble description. 
 This is the so-called solvent-controlled adiabatic ET
 which requires $V_{\rm tun}\ll k_BT,\lambda$. The relaxation 
 of populations is approximately single-exponential for activation barriers exceeding several $k_BT$,
 \begin{eqnarray}\label{population}
p_{1,2}(t)=p_{1,2}(\infty )+[p_{1,2}(0)-p_{1,2}(\infty )]e^{-kt}
\end{eqnarray}
with $p_{1,2}(\infty)=1/[1+\exp(\pm \epsilon_0/k_BT)]$, and $k=k_1+k_2$.

\begin{figure}
\vspace{1cm}
\resizebox{1.0\columnwidth}{!}{\includegraphics{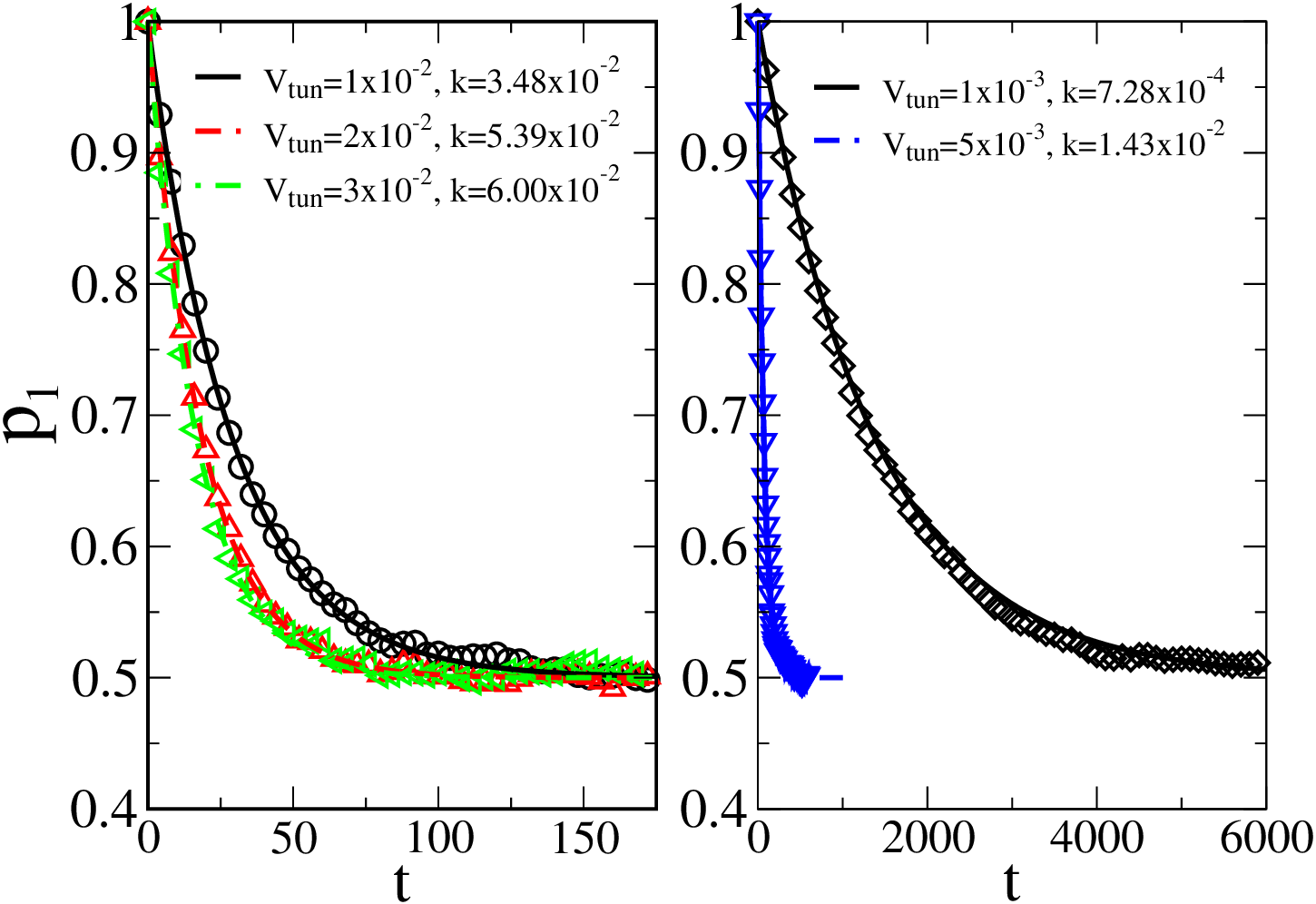}}
\caption{(Color online) Numerical relaxation of $p_1(t)$ (symbols) \textit{vs.}  the  analytical results 
in Eqs. (\ref{analytic1}), (\ref{analytic1a}), (\ref{population}) (lines) for 
$\epsilon_0=0$, $\lambda=800$, $T=0.1$ and various values of $V_{\rm tun}$ shown in 
the plot  \cite{numerics}. For $V_{\rm tun}=0.001$, ET
is nearly non-adiabatic, while for $V_{\rm tun}=0.03$ already close
to adiabatic ET with $k^{(\rm ad)}=0.066$, from the ensemble perspective.
The numerically fitted values of $k_{\rm num}$ (not shown) agree with the theoretical results
shown in the plot with the accuracy better than 0.5\% except for $V_{\rm tun}=0.001$ (about 4\%).
$N=10^4$ particles are used in simulations.
}
\label{Fig2} 
\end{figure}

In this Letter we focus on the trajectory counterpart of this well-known ensemble theory.
It can be obtained as follows. We propagate
overdamped  (with $M=0$) Langevin dynamics  (\ref{Langevin}) on one potential surface.
Once the threshold $x^*$ is reached the quantum hop on another surface occurs
with the LZS probability (\ref{LZS}), where $v=\delta x/\delta t$, $\delta t$
is the time integration step, and $\delta x$ is the $x$ displacement by crossing
the threshold. After a quantum jump, Langevin dynamics is continuously propagated on the 
other surface, on so on.  Notice that even if for  $\delta t\to 0$ the formal limit of $\delta x/\delta t$ 
 does not exist in a mean-square sense for the strictly overdamped dynamics, at any finite
$\delta t$, $v$ is finite. The overdamped dynamics of the reaction coordinate
leads, however, to an effective linearization of Eq. (\ref{LZS}) in $f(v)$, 
$P_{\rm LZ}(v)\approx f(v)$, i.e. the results do not depend on whether we 
use Eq. (\ref{LZS}), or (\ref{Landau}) in simulations. This is our first 
remarkable result which is completely
confirmed by numerics and agrees with the Zusman equations theory. 
We consider the symmetric case $\epsilon_0=0$
 in this work.
By propagating many particles simultaneously starting
from the quantum state ``1'' and distributing initial $x(0)$
in accordance with $P^{(eq)}_1(x)$, we can keep track of the state populations.
The corresponding results in Fig. \ref{Fig2} \cite{numerics} agree remarkably well with
the theoretical result in Eqs. (\ref{Marcus})-(\ref{population}). In other words, 
the ensemble
averaged trajectory result nicely agrees with the analytical solution of Zusman equations.
For a very small $V_{\rm tun}$, ET occurs non-adiabatically with the MLD rate. Upon 
increase of $V_{\rm tun}$, adiabatic
transport regime is gradually approaching. It is almost reached for $V_{\rm tun}=0.03$ in Fig. 
\ref{Fig2}. 

\begin{figure}
\vspace{1cm}
\resizebox{1.0\columnwidth}{!}{\includegraphics{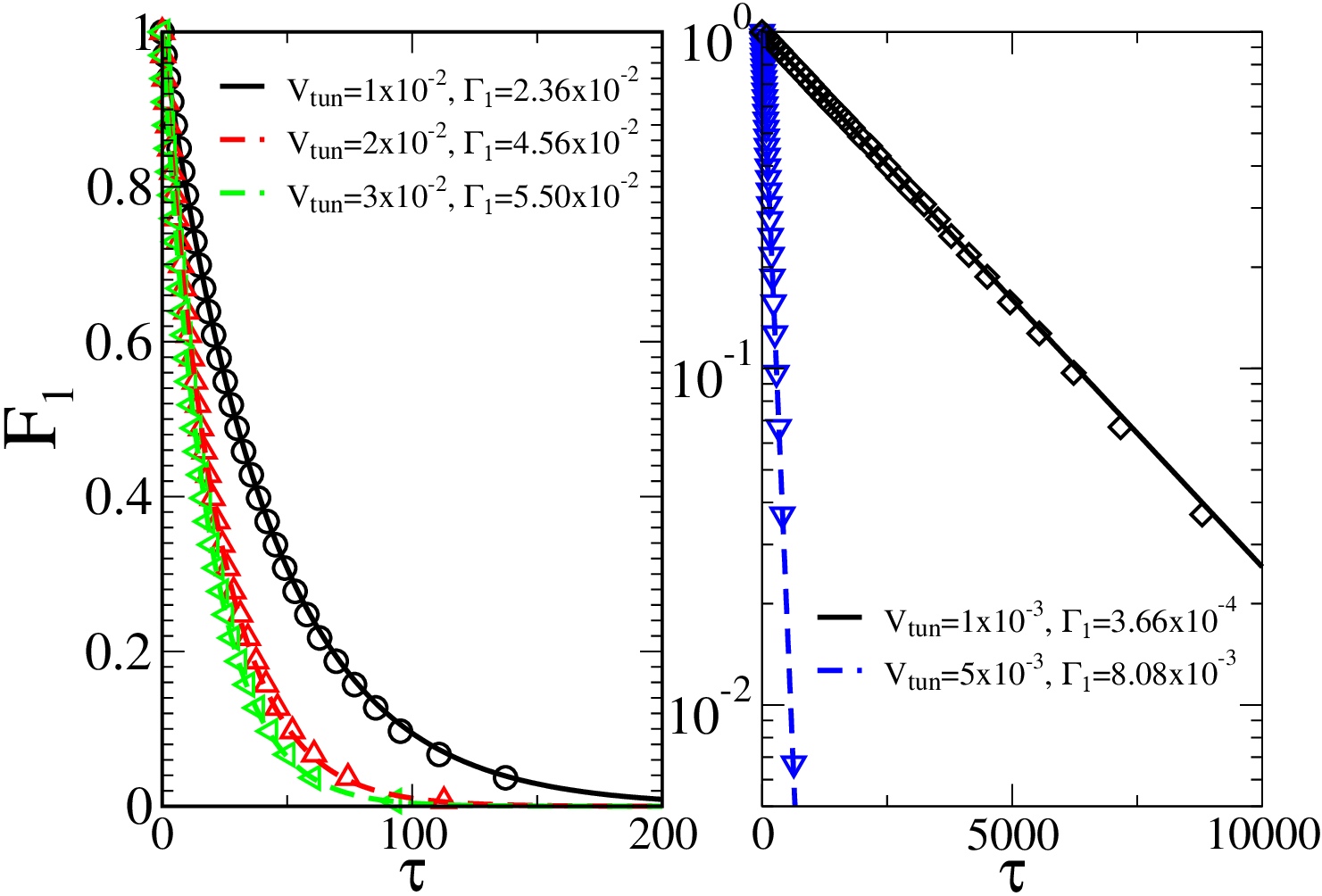}}
\caption{(Color online) Survival probability in one electronic state on the
ensemble level. Parameters are the same as in Fig. \ref{Fig2}.
Numerical results are depicted by symbols, whereas an exponential decay
with the rate in Eq. (\ref{analyt3}) is shown by lines.}
\label{Fig3} 
\end{figure}  
 
\begin{figure}
\vspace{1cm}
\resizebox{0.8\columnwidth}{!}{\includegraphics{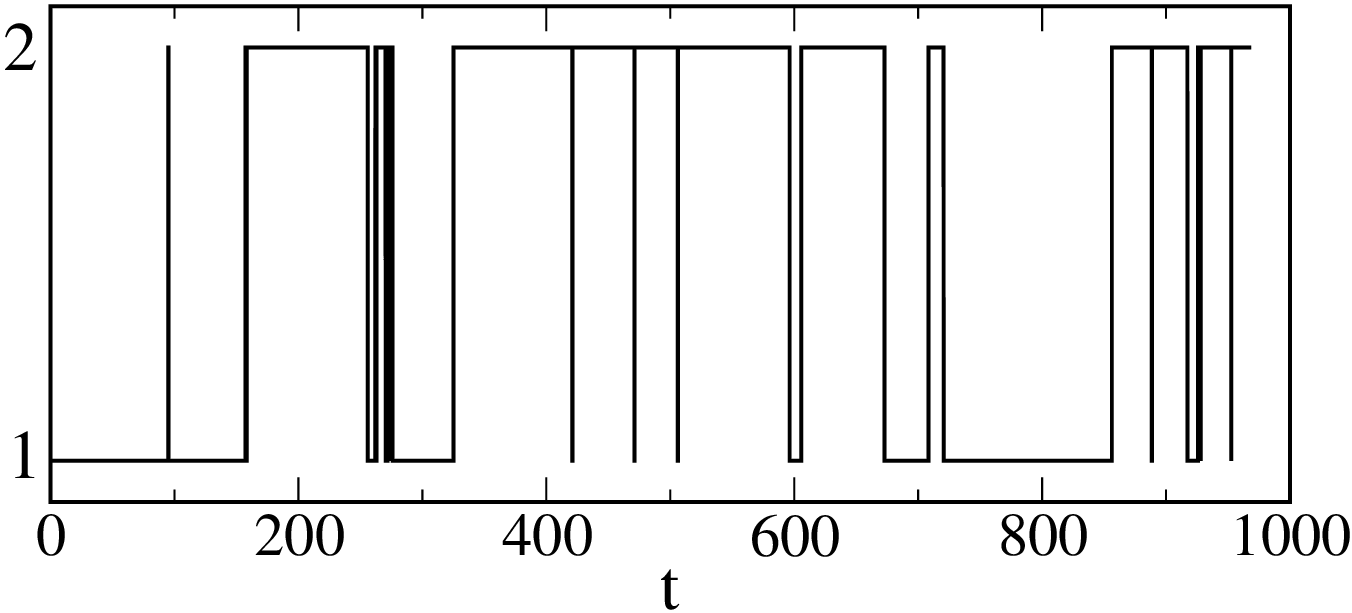}}
\caption{Quantum state trajectory realization in non-Markovian regime with broken ergodicity.
$\epsilon_0=0$, $\lambda=800$, $T=0.1$, and $V_{\rm tun}=0.01$.}
\label{Fig4} 
\end{figure}

Trajectory simulations contain, however,
much more information than Zusman equations can deliver.
 We can study also 
the residence time distributions (RTDs) in the electronic states. %Here we reveal a great surpize.
The RTD distribution on the ensemble level can be obtained by
preparing all the particles in one state, with the reaction coordinate initially
thermally equilibrated and taking out particles once they jumped  to another state until
no particles remained in the initial state. The corresponding survival probability $F_1(\tau)$
decays single-exponentially, see in Fig. \ref{Fig3}, however, with the rate $\Gamma_1$, 
which is different
from the above $k_1$. Indeed, on theoretical grounds one can maintain that
\begin{eqnarray}\label{analyt3}
\frac{1}{\Gamma_{1,2}}=\frac{1}{k_{ 1,2}^{\rm (nad)}}+\tau^{\rm (ad)}_{1,2},
\end{eqnarray}
i.e. the average time to make a transition is the sum of the average time to reach
the threshold $x^*$ and  of the inverse of the nonadiabatic tunneling rate. Indeed,
numerics remarkably agree with this statement, see in Fig. \ref{Fig3}. Furthermore, 
for a Markovian dynamics it must be $\Gamma_{1,2}=k_{1,2}$. This is indeed the case in  the
nonadiabatic ET regime characterized by MLD rate. 
 However, dynamics
of electronic transitions becomes increasingly  non-Markovian upon taking  adiabatic
corrections into account with the increase of $V_{\rm tun}$. This is 
in spite of a single-exponential
character of the ET kinetics on the ensemble level! 
Ref. \cite{GoychukJCP05} already pointed out on a similar very paradoxical situation:
a highly non-Markovian bursting process can have a nearly exponentially decaying autocorrelation
function.  Indeed, a short inspection of
a single trajectory realization of electronic transitions in 
such a non-Markovian regime depicted in Fig. \ref{Fig4} reveals immediately its
non-Markovian character. Bursting provides a visual proof \cite{GoychukJCP05}.
Notice that a popular statement that in adiabatic ET regime electrons just
follow to nuclear transitions is in fact very misleading on the level of single electron
trajectories. This is so because electron jumps only at the level
crossings (in the contact approximation) and the ensemble description on the level of populations
relaxation completely misses this very essential quantum mechanical feature.
ET remains quantum even within this adiabatic seemingly fully classical regime!
And namely this causes a \textit{quantum} breaking of ergodicity discovered
next.

\begin{figure}
\vspace{1cm}
\resizebox{0.8\columnwidth}{!}{\includegraphics{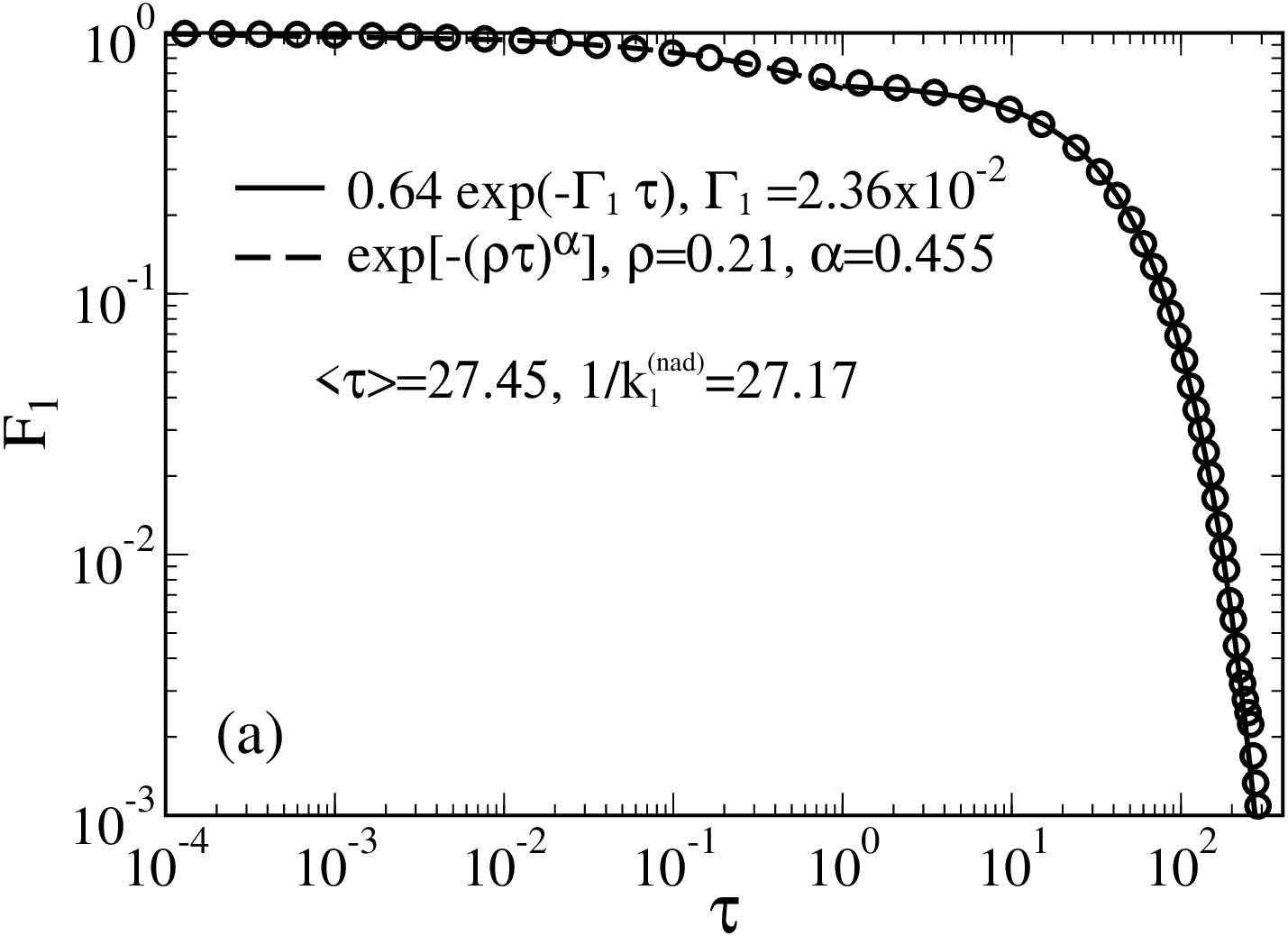}}\\
\vfill \vspace{.7cm}
\resizebox{0.8\columnwidth}{!}{\includegraphics{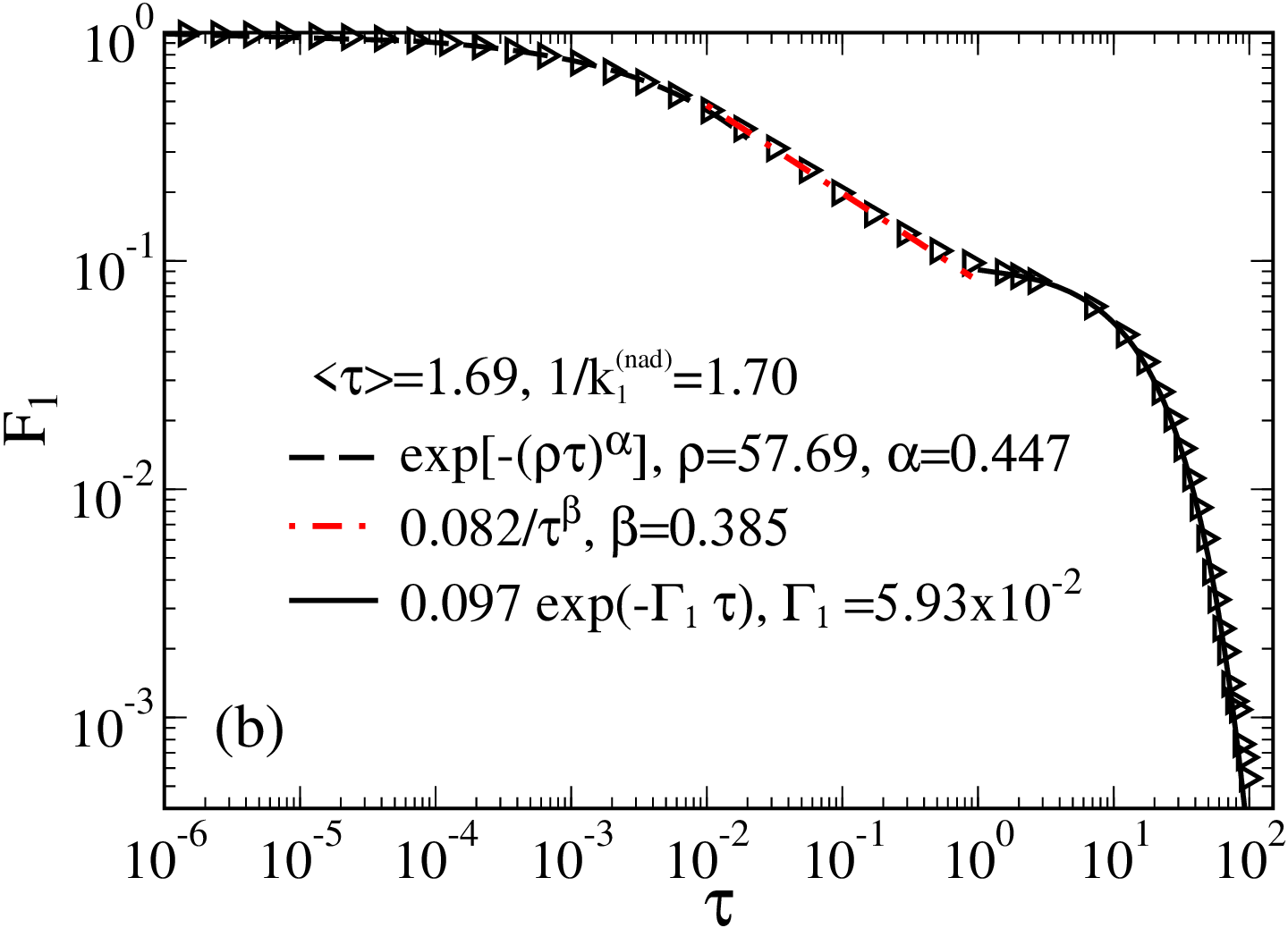}}
\caption{(Color online) Survival probability in one electronic state on  a
single trajectory level. Parameters are the same as in Fig. \ref{Fig2},
and (a) $V_{\rm tun}=0.01$ (as in Fig. \ref{Fig4}),  (b) $V_{\rm tun}=0.04$.
Numerical results are depicted by symbols and their fits by lines detailed in the plots. }
\label{Fig5} 
\end{figure}   

Indeed, the study of survival probabilities based on single very long trajectories reveals a real
surprise indicating breaking of ergodicity in this profoundly non-Markovian regime.
The corresponding survival probability in a state is depicted
in Fig. \ref{Fig5}, (a).
It is profoundly non-exponential, very differently from the corresponding ensemble result 
in Fig. \ref{Fig3}. The rate $\Gamma_1$ describes only the tail of distribution,
which is initially stretched exponential. It can possess also an intermediate 
power law regime for a larger $V_{\rm tun}$, see part (b) in Fig. \ref{Fig5}, where
the exponential tail has weight less than 10\%.
Very surprisingly, the mean residence time is well described by the inverse of 
the Marcus-Levich-Dogonadze rate, $\langle \tau_i\rangle=1/k_i^{\rm (nad)}$.
This can be explained within a modification of the classical
level-crossing theory \cite{Papoulis}. Let us take formally into account small
inertial effects (keeping first $M$ finite). Then, the process $v(t)$ is not singular.
Consider dynamics in the state $i$. Assuming stationarity of $x(t)$, the averaged number 
of level crossings $n_i({\cal T})$ within a very long time interval ${\cal T}$ is \cite{Papoulis} 
$n_i({\cal T})={\cal T}P_i^{(eq)}(x^*)\langle |v(t)|\rangle_{x(t)=x^*}$, and hence
$\langle \tau_i\rangle^{-1}=\lim_{{\cal T}\to \infty}n_i({\cal T})/{\cal T}=
P_i^{(eq)}(x^*)\langle |v(t)|\rangle_{x(t)=x^*}$. By the same token
and taking into account the probability (\ref{LZS}) to make a quantum jump to another state
at each level
crossing we obtain 
\begin{eqnarray}\label{gen}
\langle \tau_i\rangle^{-1}=P_i^{(eq)}(x^*)\langle |v| P_{\rm LZ}(v)\rangle_{x(t)=x^*}\;\;.
\end{eqnarray}
%for the inverse mean residence time in this state.
Averaging in (\ref{gen}) with Maxwellian equilibrium $P_M(v)$ yields a very important result
\begin{eqnarray}\label{new1}
\langle \tau_i\rangle^{-1}=k_i^{\rm (nad)}R(z=v_0/v_T)\;,
\end{eqnarray} 
where
\begin{eqnarray} \label{new2}
R(z)=\sqrt{\frac{2}{\pi z^2}}-\frac{1}{2\pi}
\MeijerG[\Bigg]{3}{0}{0}{3}{-}{\frac{1}{2},0, -\frac{1}{2}}{\frac{z^2}{8}}\quad
\end{eqnarray}
is a renormalization function taking inertial effects into account. It is expressed
via a Meijer G-function \cite{Gradstein}, and $v_0=\pi |V_{\rm tun}|^2x_0/(\hbar\lambda)$ is
a characteristic tunnel velocity. Numerically, 
$R(z)\approx \exp(-1.57z^{0.9})$ for $0<z<0.1$ with the 
accuracy of about 10\%. In the formal overdamped limit, 
$\lim_{M\to 0} R(v_0/v_T)=1$, and we obtain 
$\langle \tau_1\rangle^{-1}=k_1^{\rm (nad)}$, in agreement with numerics. 
Moreover,
we did also numerics which include inertial effects in Eq. (\ref{Langevin}) and confirm the analytical
result in (\ref{new1}), (\ref{new2}) \cite{remark}.  The observed
ergodicity breaking is thus not an artifact of the overdamped singular approximation.
It expresses quantum nature of electron transfer even in adiabatic regime as
manifested on the level of single molecule dynamics.

As a major result of this work, equations like Zusman equations
and other quantum ensemble descriptions simply cannot be used to describe properties
of profoundly non-Markovian  single electron trajectories.  This can be relevant e.g.
for blinking quantum dots
in non-exponential regimes, whenever the reaction coordinate dynamics is very
essential  \cite{TangMarcus}. This is especially true for anomalously slow subdiffusive 
dynamics which is the subject of a separate follow-up work. The
discovered ergodicity breaking in a simple and well-known model of charge transport dynamics
is expected to influence a large body of current research.

\textit{Acknowledgment.} 
Support of this research by the Deutsche Forschungsgemeinschaft (German Research Foundation), Grant
GO 2052/1-2 is gratefully acknowledged.


\begin{references}
\bibitem{remark} A detailed treatment will be presented elsewhere.
\end{references}
\end{document}